# BiOBr 2D materials for integrated nonlinear photonics devices

Linnan Jia[a], Dandan Cui[b], Jiayang Wu[a], Haifeng Feng[c], Yunyi Yang[d],
Tieshan Yang[d], Yang Qu[a], Yi Du[c], Weichang Hao[b], Baohua Jia[a,d], and David J. Moss[a*]

[a]Centre for Micro-Photonics, Swinburne University of Technology, Hawthorn, VIC 3122, Australia;

[b]School of Physics, and BUAA-UOW Joint Research Centre, Beihang University, Beijing 100191, China;

[c]Institute for Superconducting and Electronic Materials, and UOW-BUAA Joint Research Centre, University of Wollongong, Wollongong, NSW 2500, Australia;

[d]Centre for Translational Atomaterials, Swinburne University of Technology, Hawthorn, VIC 3122, Australia

* Electronic mail: dmoss@swin.edu.au

## ABSTRACT

As a new group of advanced 2D layered materials, bismuth oxyhalides, i.e., BiOX (X = Cl, Br, I), have recently become of great interest. In this work, we characterize the third-order optical nonlinearities of BiOBr, an important member of the BiOX family. The nonlinear absorption and Kerr nonlinearity of BiOBr nanoflakes at both 800 nm and 1550 nm are characterized via the Z-Scan technique. Experimental results show that BiOBr nanoflakes exhibit a large nonlinear absorption coefficient $\beta \sim 10^{-7}$ m/W as well as a large Kerr coefficient $n_2 \sim 10^{-14}$ m$^2$/W. We also note that the $n_2$ of BiOBr reverses sign from negative to positive as the wavelength is changed from 800 nm to 1550 nm. We further characterize the thickness-dependent nonlinear optical properties of BiOBr nanoflakes, finding that the magnitudes of $\beta$ and $n_2$ increase with decreasing thickness of the BiOBr nanoflakes. Finally, we integrate BiOBr nanoflakes into silicon integrated waveguides and measure their insertion loss, with the extracted waveguide propagation loss showing good agreement with mode simulations based on ellipsometry measurements. These results confirm the strong potential of BiOBr as a promising nonlinear optical material for high-performance hybrid integrated photonic devices.

## 1. INTRODUCTION

Featuring broad operation bandwidths, low power consumption, and potentially reduced cost, all-optical signal processing based on nonlinear photonic devices has provided a competitive solution to realize ultrafast signal processing.[1-7] As the key building blocks for implementing nonlinear photonic devices, advanced optical materials with superior nonlinear properties have been widely investigated. Recently, two-dimensional (2D) layered materials such as graphene[8, 9], graphene oxide (GO) [10-14], transition metal dichalcogenides (TMDCs) [15-17], and black phosphorus (BP) [18-20] have attracted significant interest with their remarkable optical properties, such as ultrahigh Kerr optical nonlinearities, strong nonlinear absorption, significant material anisotropy, and layer-dependent material properties. These 2D materials have already enabled various new photonic devices which are fundamentally different from those based on traditional bulk materials [21-23].

In addition to those typical 2D materials, Bismuth oxyhalides, i.e., BiOX (X = Cl, Br, I), have been explored as a new group of advanced 2D layered materials [24-26]. BiOX have an open-layer crystal structure which is built by interlacing [Bi$_2$O$_2$] slabs with double halogen slabs. This unique structure enables self-built internal static electric fields in BiOX, which lead to an effective separation of photoinduced charge carriers, thus enabling BiOX with superior photocatalytic behavior [26-28] and excellent nonlinear optical performance [29, 30]. The strong nonlinear optical properties of BiOCl, in terms of both nonlinear absorption and Kerr nonlinearity, have been demonstrated at 515 nm[29] and 800 nm[30].

In this work, we investigate the nonlinear optical properties of BiOBr nanoflakes. The nonlinear absorption and Kerr nonlinearity of BiOBr nanoflakes at both 800 nm and 1550 nm are characterized via the Z-Scan technique [31]. We demonstrate BiOBr nanoflakes have strong nonlinear absorption and high Kerr nonlinearity at both wavelengths, with a large nonlinear absorption coefficient $\beta$ of $\sim 10^{-7}$ m/W as well as a large Kerr coefficient $n_2 \sim 10^{-14}$ m$^2$/W. A strong dispersion of Kerr coefficient $n_2$ of BiOBr is also observed in our experiment, where $n_2$ reverses sign from negative to positive as the wavelength is changed from 800 nm to 1550 nm. In addition, thickness dependent nonlinear response of the BiOBr nanoflakes are verified, with $\beta$ and $n_2$ increasing for very thin flake thicknesses. We also integrate the BiOBr nanoflakes onto silicon integrated waveguides and measure the insertion loss of the hybrid integrated devices, with the extracted waveguide propagation loss showing good agreement with mode simulations based on ellipsometry measurements. Our results verify the strong potential of BiOBr as an advanced optical material for high-performance nonlinear photonic devices.

## 2. MATERIAL PREPARATION AND CHARACTERIZATION

Fig. 1(a) illustrates the atomic structure of BiOBr crystals, where $[Bi_2O_2]^{2+}$ slabs are interleaved by double Br atoms to form a layered structure [24, 26]. The thickness profiles of the prepared BiOBr nanoflakes were characterized by atomic force microscopy (AFM), as shown in Fig. 1(b). The measured thicknesses of the samples in (i) − (iv) are ~30 nm, ~75 nm, ~110 nm, and ~140 nm, respectively.

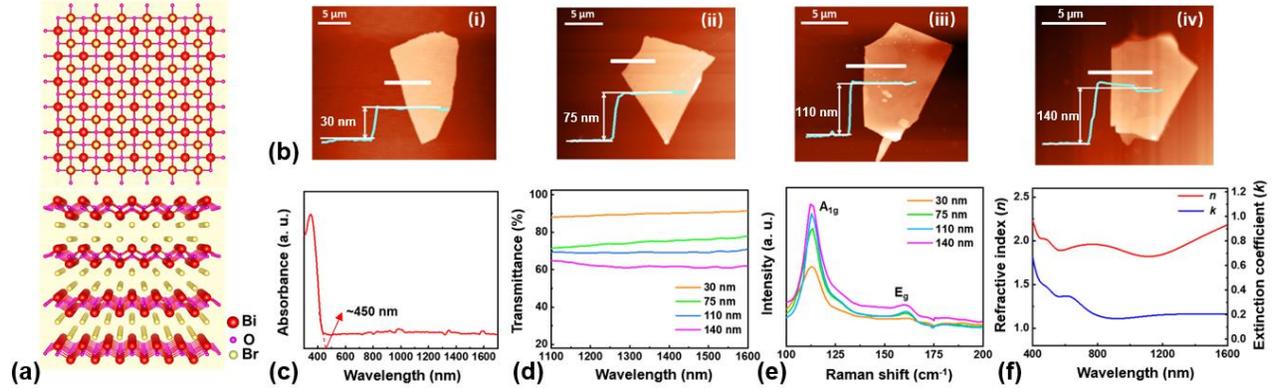

Fig. 1. (a) Schematic atomic structure of BiOBr. (b) AFM images and height profiles of exfoliated BiOBr nanoflakes with various thicknesses: (i) ~30 nm, (ii) ~75 nm, (iii) ~110 nm, (iv) ~140 nm. (c) UV-vis absorption spectrum of BiOBr. (d) Measured linear transmittance spectra of BiOBr nanoflakes with different thicknesses. (e) Raman spectra of BiOBr nanoflakes with different thicknesses. (f) Measured refractive index ($n$) and extinction coefficient ($k$) of BiOBr.

Fig. 1(c) depicts the linear absorption of BiOBr from 300 nm to 1700 nm measured by ultraviolet-visible (UV-vis) spectrometry. A strong linear absorption edge near ~ 450 nm is observed, which corresponds to a photon energy of ~ 2.76 eV, in agreement with the reported bandgap of BiOBr [28]. We also measured the transmittance spectra of BiOBr nanoflakes with different thicknesses, as shown in Fig. 1(d). High transmittance (> 60%) for wavelengths from 1100 nm to 1600 nm is observed for all samples, with a transmittance reaching 90% for the 30-nm-thick BiOBr nanoflakes. To characterize the quality of prepared BiOBr samples, Raman spectroscopic measurements were conducted with a pump laser wavelength of ~532 nm. As shown in Fig. 1 (e), two phonon modes of $A_{1g}$ (~113.2 cm$^{-1}$) and $E_g$ (~160.4 cm$^{-1}$) are observed for all samples, which is consistent with previous reports [32, 33]. Fig. 1(f) shows the in-plane refractive index ($n$) as well as extinction coefficient ($k$) of BiOBr measured by spectral ellipsometry [11, 34]. The sample thickness is ~ 1 µm. The measured $n$ and $k$ in telecommunications band are ~2.2 and ~ 0.2, respectively.

## 3. Z-SCAN MEASUREMENT

We characterize the nonlinear absorption and refraction of the prepared BiOBr samples via Z-scan techniques. [20, 31, 35] The experimental setup is illustrated in Fig. 2. Femtosecond pulsed lasers, with center wavelengths at ~800 nm and ~1550 nm, were used to excite the samples. The repetition rate and pulse duration were ~80 MHz and ~140 fs, respectively. The incident laser beam was focused by an objective lens (10×, 0.25NA) to achieve a low beam waist with a focused spot size much smaller than the sample size, at ~1.6 µm and ~3.1 µm for the 800-nm and 1550-nm pulsed lasers, respectively.

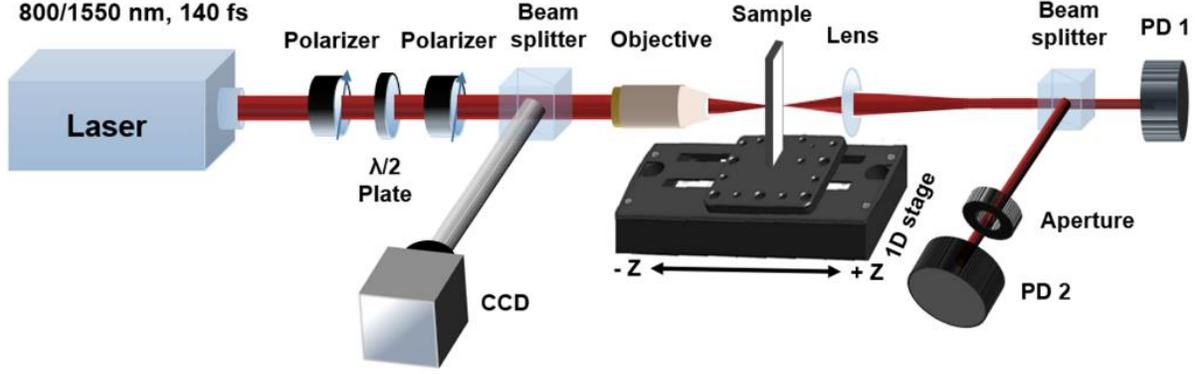

Fig. 2. Schemes illustration of Z-scan experimental setup. PD: power detector.

Fig. 3(a) and (b) depict the Z-scan results of a BiOBr nanoflake (~140 nm) at 800 nm and 1550 nm, respectively with (i) showing the OA results and (ii) the CA results. In the OA measurement (Figs. 3(a-i) and 3(b-i)), we observe the typical reverse saturation absorption (RSA) [16, 18, 29] at both wavelengths. Since the optical bandgap of BiOBr is much larger than the photo energy of incident lasers, we attribute the observed RSA to the two-photon absorption (TPA) at 800 nm and multi-photon absorption (MPA) at 1550 nm. To extract the nonlinear coefficient $\beta$ of BiOBr, we fit the measured OA data by [19, 36]:

$$T_{OA}(z) \simeq 1 - \frac{1}{2\sqrt{2}} \frac{\beta I_0 L_{eff}}{(1+x^2)}, \tag{1}$$

where $T_{OA}(z)$ is the normalized optical transmittance of OA measurement, and $x = z/z_0$, with $z$ and $z_0$ denoting the sample position relative to the focus and the Rayleigh length of the laser beam, respectively; $L_{eff} = (1-e^{-\alpha_0 L})/\alpha_0$ is the effective sample thickness, with $\alpha_0$ and $L$ denoting the linear absorption coefficient and the sample thickness, respectively and $I_0$ is the irradiance intensity at focus. The $\beta$'s at 800 nm (Fig. 3(a-i)) and 1550 nm (Fig. 3(b-i)) are measured to be ~$1.869 \times 10^{-7}$ m/W and ~$1.554 \times 10^{-7}$ m/W, respectively. Such high nonlinear absorption $\beta$ is very favorable for high performance optical limiter applications.[8, 10]

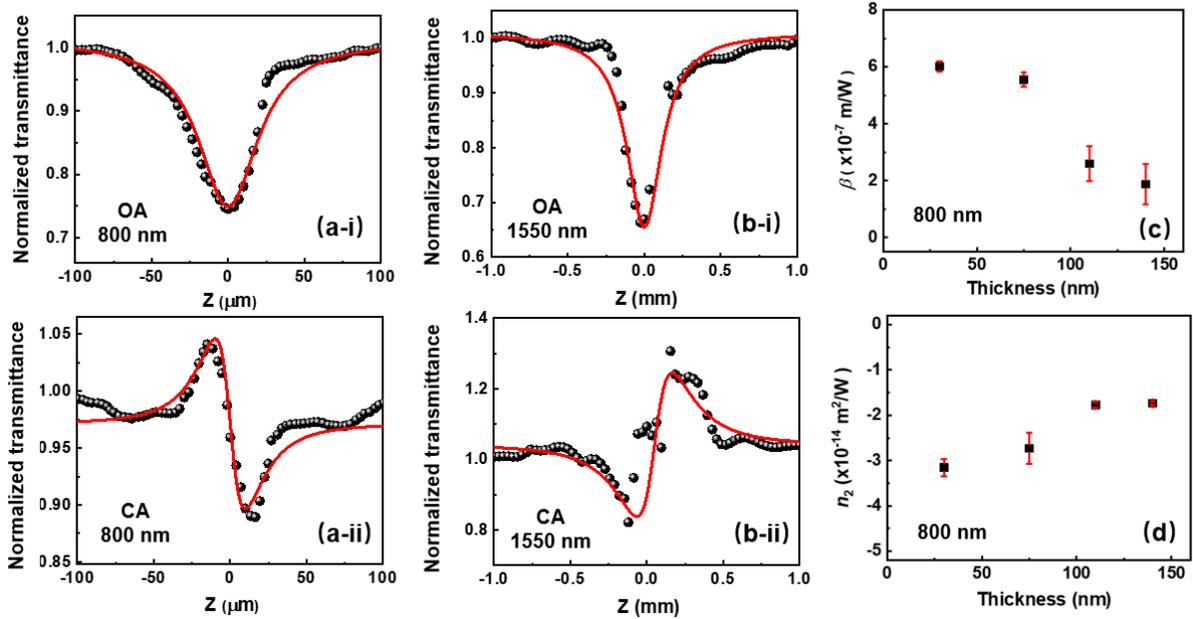

Fig. 3. (a) OA (i) and CA (ii) results at 800 nm. (b) OA (i) and CA (ii) results at 1550 nm. The thickness of the BiOBr sample is ~140 nm. (c) nonlinear absorption coefficient $\beta$ and (d) Kerr coefficient $n_2$ for BiOBr samples with different thicknesses.

We also investigate the nonlinear refraction of the BiOBr sample at both 800 nm and 1550 nm with the CA measurements, as shown in Figs. 3(a-ii) and (b-ii). In Fig. 3(a-ii), prominent peak-valley transmittance is observed for the CA measurement at 800 nm, which corresponds to a negative Kerr coefficient $n_2$ and reflects optical self-defocusing in BiOBr nanoflakes. In contrast, valley-peak transmittance is observed at 1550 nm in Fig. 3(b-ii), which is a reflection of optical self-focusing, resulting in a positive $n_2$. The $n_2$ value of BiOBr can be extracted from the measured CA results with [31]:

$$T_{CA}(z, \Delta\Phi_0) \simeq 1 + \frac{4\Delta\Phi_0 x}{(x^2+9)(x^2+1)}, \qquad (2)$$

where $T_{CA}(z, \Delta\Phi_0)$ is the normalized optical transmittance of CA measurement. $\Delta\Phi_0 = 2\pi n_2 I_0 L_{eff}/\lambda$ is the nonlinear phase shift, with $\lambda$ denoting the laser center wavelength. The measured Kerr coefficients $n_2$'s at 800 nm and 1550 nm are $\sim$-1.737 $\times 10^{-14}$ m$^2$/W and $\sim$3.824 $\times 10^{-14}$ m$^2$/W, respectively. The transition from negative $n_2$ at 800 nm to positive $n_2$ at 1550 nm can be attributed to the dispersion of $n_2$ associated with the two-photon bandgap where $n_2$ is positive when the excitation photon energy (at 1550 nm) is below the TPA bandedge (half-bandgap) while it becomes negative when the photon energy (at 800 nm) is between the one-photo absorption and TPA edges [37, 38]. It should be note that the $n_2$ value of BiOBr is more than three orders of magnitude larger than that of Si[1], demonstrating its prominent Kerr nonlinearities and high potential for high-performance nonlinear photonic applications.

We further characterize the thickness dependent nonlinear response of BiOBr nanoflakes via 800 nm Z-scan measurements. The irradiance laser intensity was 0.202 GW/cm$^2$. The measured $\beta$ and $n_2$ for BiOBr samples with different thicknesses are plotted in Figs. 3(c) and (d), respectively. It can be seen that $\beta$ and $n_2$ for the BiOBr nanoflakes is highly thickness dependent, where their absolute values increase when the sample thickness decreases from 140nm to 30 nm. The dependence of nonlinear optical parameters on sample thickness is likely induced by localized defects in BiOBr nanoflakes, which would lead to more scattering and energy loss for thicker BiOBr nanoflakes, thus resulting in a decreased absolute values of $\beta$ and $n_2$ [36, 39].

**Table 1.** Comparison of $\beta$, $n_2$, and figure of merit (FOM) of various 2D layered materials

| Material | Laser parameter | Thickness (nm) | $\beta$ (m/W) | $n_2$ (m$^2$/W) | FOM | Ref. |
|---|---|---|---|---|---|---|
| Graphene | 1550 nm, 100 fs | 5-7 layers | $9 \times 10^{-8}$ | $-8 \times 10^{-14}$ | -0.574 | [9] |
| GO | 800 nm, 100 fs | $2 \times 10^3$ | $4 \times 10^{-7}$ | $1.25 \times 10^{-13}$ | 0.391 | [10] |
| MoS$_2$ | 1064 nm, 25 ps | $2.5 \times 10^4$ | $-3.8 \times 10^{-11}$ | $1.88 \times 10^{-16}$ | -4.649 | [15] |
| WS$_2$ | 1040 nm, 340 fs | 57.9 | $1.81 \times 10^{-8}$ | $-3.36 \times 10^{-16}$ | -0.018 | [16] |
| BP | 1030 nm, 140 fs | 15 | $5.845 \times 10^{-6}$ | $-1.635 \times 10^{-12}$ | -0.272 | [19] |
| BiOCl | 800 nm, 100 fs | 20-140 | $4.25 \times 10^{-9}$ | $3.8 \times 10^{-15}$ | 1.118 | [30] |
| BiOBr | 800 nm, 140 fs | 30 | $6.011 \times 10^{-7}$ | $-3.155 \times 10^{-14}$ | -0.066 | This work |
| BiOBr | 800 nm, 140 fs | 140 | $1.869 \times 10^{-7}$ | $-1.737 \times 10^{-14}$ | -0.116 | This work |
| BiOBr | 1550 nm, 140 fs | 140 | $1.554 \times 10^{-7}$ | $3.824 \times 10^{-14}$ | 0.159 | This work |

In Table 1, we compare the measured $\beta$ and $n_2$ of BiOBr with other 2D layered materials. What is significant is that, BiOBr nanoflakes exhibit a large $\beta$ on the order of $10^{-7}$ m/W, which is much higher than many other layered materials, in particular being two orders of magnitude higher than BiOCl. The Kerr coefficient $n_2$ is on the order of $10^{-14}$ m$^2$/W, which is close to that of graphene and GO, and is more than one order of magnitude higher than TMDCs and BiOCl. These results confirm the superior nonlinear optical properties of BiOBr nanoflakes as an advanced nonlinear optical material.

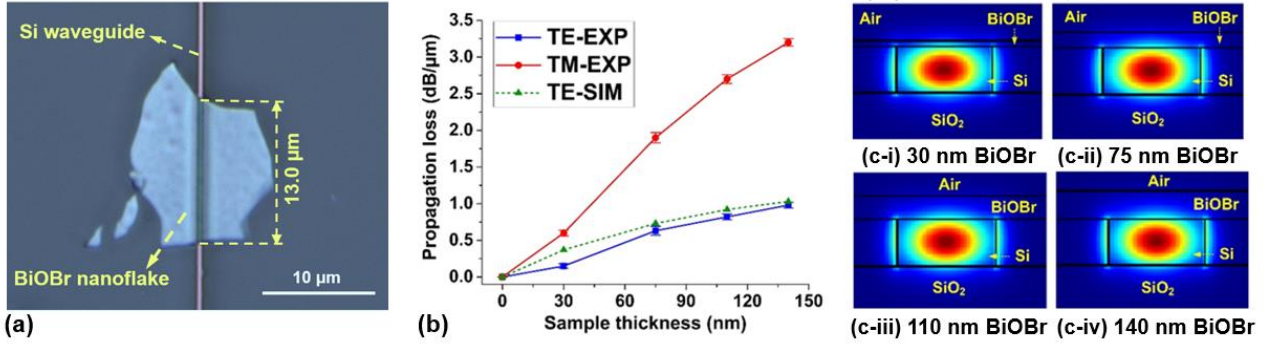

Fig. 4 (a) Microscope image of a silicon integrated waveguides incorporated with BiOBr nanoflake. (b) Measured and simulated waveguide propagation loss of the hybrid waveguides for different BiOBr thicknesses. (c) TE (Ex) mode profile of the hybrid integrated waveguide for different BiOBr thicknesses: (i) 30 nm, (ii) 75 nm, (iii) 110 nm, and (iv) 140 nm.

## 4. INTEGRATION ON SILICON PHOTONIC DEVICES

Finally, we characterize the BiOBr nanoflakes integrated in 220-nm-thick silicon-on-insulator (SOI) waveguides [40] [1] on a 2-μm-thick buried oxide (BOX) layer. BiOBr nanoflakes were transferred onto the silicon integrated waveguides using an all-dry transfer method [36, 41]. Fig. 4(a) shows a representative microscope image of a silicon integrated waveguide incorporated with BiOBr nanoflake. The width of the waveguide was ~500 nm. It can be seen that the BiOBr nanoflake is attached to the silicon integrated waveguide, with an overlap length of ~13 μm. The thickness of the BiOBr nanoflake is ~110 nm.

The measured TE polarized insertion losses of the silicon waveguides with and without a 110-nm-thick BiOBr nanoflake are ~6.9 dB and ~16.0 dB, respectively. When we measured the insertion losses, the power of the input continuous-wave light was ~0 dBm. The butt coupling loss was ~3.0 dB each, or ~6.0 dB for both. According to our previously fabricated devices, the waveguide propagation loss for single-mode silicon nanowire waveguides with a cross section of 500 nm × 220 nm is about ~3 dB/cm [42, 43], which is much lower than that of the hybrid waveguide. Therefore, the propagation loss of the hybrid waveguide can be given by:

$$PL = (IL_{hybrid} - IL_{silicon}) / L_{hybrid}, \qquad (3)$$

where $IL_{hybrid}$ and $IL_{silicon}$ are the insertion losses of the hybrid waveguide and the silicon waveguide without BiOBr, respectively. $L_{hybrid}$ is the length of the BiOBr nanoflake on the silicon waveguide. The TE and TM polarized waveguide propagation losses extracted from the measured insertion losses is shown in Fig. 4(b). We measured the hybrid waveguides with four different BiOBr thicknesses. The data points depict the average values obtained from the experimental results of three samples and the error bars illustrate the variations for different samples. The propagation loss for TM polarization is much higher than that for TE polarization. Such a difference is mainly caused by mode overlap and can be used for implementing polarizers [44]. We also perform mode analysis for the hybrid integrated waveguide using Lumerical FDTD commercial mode solving software. We used the in-plane (TE-polarized) $n$ and $k$ of 1-μm-thick BiOBr obtained from the ellipsometry measurements (the values in Fig. 1(f) at 1550 nm) in the FDTD simulation. Fig. 4(c) shows the TE mode profile of the hybrid integrated waveguides with different BiOBr thicknesses. For comparison, the simulated waveguide propagation losses are also shown in Fig. 4(b). It can be seen that the experimental propagation losses are close to the simulated propagation losses, which reflects the stability of the prepared BiOBr nanoflakes. We also note that the simulated propagation losses based on the $n$ and $k$ of 1-μm-thick BiOBr are slightly higher than the experimental propagation losses, with an increased difference between them for a decreased BiOBr thickness. This indicates that the intrinsic material loss actually increases with thickness, which could be attributed to any number of effects such as increased scattering loss and absorption induced by imperfect contact between the multiple layers as well as interactions between them.

One of the most promising applications for highly nonlinear 2D materials is for photonic chips operating all-optically via the 3rd order optical nonlinearity. All-optical signal processing based on silicon [45,46 28, 29] has been extremely successful for all-optical logic [47 30], ultra-high speed demultiplexing [48,49 31, 32], optical performance monitoring [50,51 33, 34], regeneration [52,53 35, 36], and others [54-60 37-43]. Since CMOS (complementary metal oxide semiconductor) compatible platforms are centrosymmetric, nonlinear devices have been based on third order nonlinearities - the Kerr nonlinearity ($n_2$) [45,46 28, 29] and third harmonic generation [55, 61-65 38, 44-48]. However, while silicon has an extremely high nonlinearity γ, it also has high two-photon absorption (TPA, β) and a poor nonlinear figure of merit of 0.3 (FOM = $n_2$ / (β λ)) in the telecom band. While TPA can be advantageous [66-68 49-51], it is generally a limitation and this inspired interest in other platforms including chalcogenide glasses [69-78 52-61]. In 2008 new nonlinear CMOS platforms [79-91 62-74] enabled the first integrated micro-combs [80,81 63, 64] following the discovery of Kerr combs in 2007 [92 75]. Many breakthroughs have since been reported including mode-locked lasers [93-96 76-79], quantum physics [97-103 80-86], optical frequency synthesis [104 87], ultrahigh bandwidth communications [105 88], and others [106-112 89-95]. The success of these platforms and others such as a-Si [113 96] arises from their low linear loss, high nonlinearity, and very low TPA. BiOBr is a promising 2D material that may be of potential interest for nonlinear optical integrated circuits.

## 5. CONCLUSION

In summary, we report on the nonlinear optical properties of BiOBr nanoflakes. A large nonlinear absorption coefficient β on the order of $10^{-7}$ m/W as well as a large Kerr coefficient $n_2$ on the order of $10^{-14}$ m$^2$/W in BiOBr are observed in both 800 nm and 1550 nm Z-scan measurement. We also observe a strong dispersion in $n_2$, with $n_2$ changing sign from negative to positive when the wavelength is varied from 800 nm to 1550 nm. The thickness-dependent nonlinear optical properties of BiOBr nanoflakes are demonstrated, where the magnitudes of β and $n_2$ increase with decreasing thickness of the BiOBr nanoflakes. Finally, we integrate BiOBr nanoflakes into silicon integrated waveguides and measure their insertion loss, with the extracted waveguide propagation loss showing good agreement with mode simulations.

## ACKNOWLEDGEMENTS

This work was supported by the Australian Research Council Discovery Projects Program (No. DP150102972 and DP190103186). We acknowledge Swinburne Nano Lab and Micro Nano Research Facility (MNRF) of RMIT University for the support in material characterization as well as Advanced Micro Foundry (AMF) Pte Ltd for the support in silicon device fabrication.